\documentclass[preprintnumbers,amsmath,amssymb,prc,twocolumn]{revtex4}
\usepackage{graphicx}
\usepackage{amssymb}
\usepackage{amsmath}
\usepackage{ulem}
\usepackage{xcolor,etoolbox}
\usepackage{dcolumn}% Align table columns on decimal point
\usepackage{bm}
\begin{document}

\title{A puzzle on isomeric configurations in and around $N=126$ closed shell}

\author{Bhoomika Maheshwari$^1$, Deepika Choudhury$^{1}$, Ashok Kumar Jain$^2$}
\affiliation{$^1$Department of Physics, Indian Institute of Technology Ropar, Rupnagar 140001, India}
\affiliation{$^2$Amity Institute of Nuclear Science and Technology, Amity University UP, Noida-201313, India}

\begin{abstract}
%\lipsum[1]
The puzzle of finding consistent nuclear configurations for explaining both the decay probabilities and moments of the ${9/2}^-$, ${8}^+$ and ${21/2}^-$ isomers in and around $N=126$ closed shell has been approached in the generalized seniority scheme. Though $h_{9/2}$ is the dominant orbital near Fermi energy, the role of configuration mixing from the surrounding $f_{7/2}$ and $i_{13/2}$ orbitals is found to be very important for the consistent explanation of all the isomeric properties such as the $B(E2)$ rates, $Q-$moments and $g-$factors. The structural behavior of the closed shell $N=126$ isotonic isomers turns out to be very similar to the $N=124$ and $N=128$ isotonic isomers which have two neutron-holes and two neutron-particles, respectively. This is due to the pairing symmetries of nuclear many-body Hamiltonian. As a confirmation, the microscopic shell model occupancies are also calculated for these isomers in $N=126$ chain which support the generalized seniority results. Additional arguments using the systematics of odd-proton ${9/2}^-$ states in Tl ($Z=81$), Bi ($Z=83$), At ($Z=85$) and Fr ($Z=87$) isotopes are also presented. 
\end{abstract}
\date{\today}

\maketitle
\section{Introduction}

Isomers are the long-lived excited states of nuclei, and are of fundamental as well as industrial interest, particularly in medicine and energy~\cite{walker1999,walker2021,jain2021}. One of the important isomeric regions in nuclear chart belongs to $^{208}$Pb and neighborhood~\cite{jain2015}, which is also known for shape coexistence~\cite{heyde2011}. Most of the isomeric states in this mass region exist because of the availability of many high-$j$ intruder orbitals leading to hindered transitions. Due to the symmetries of shell model corresponding to the pairing correlations, a few isomeric states occur quite regularly in spite of the complex nuclear many-body Hamiltonian, and are known as seniority and generalized seniority isomers, mostly found in the semi-magic (spherical or nearly spherical) nuclei~\cite{frank2009,casten1990,talmi1993,shalit1963,heyde1990,isacker2014}. In order to map the boundaries where the generalized seniority (nearly spherical) regime changes to the region characterized by collective configurations~\cite{ressler2005}, it is important to study the isomeric decay properties and moments to obtain the information on nucleonic structure and underlying configurations.

Recently, we have studied the generalized seniority $v=1$, ${13/2}^+$, $v=2$, ${12}^+$, and $v=3$, ${33/2}^+$ isomers in Hg, Pb and Po isotopic chains and understood their isomeric decays and moments on the basis of a consistent multi-j neutron $i_{13/2} \otimes f_{7/2} \otimes p_{3/2} $ configuration in contrast to the common understanding of pure $i_{13/2}$ configuration~\cite{maheshwari2021}. One may then ask if similar isomers exist in various isotonic chains with protons as valence particles. The nearest neighboring $N=126$ isotonic chain would be a good example to look for such isomers. For example, the ${9/2}^-, {8}^+$, and ${21/2}^-$ isomers in $N=126$ chain are believed to be arising from proton-$h_{9/2}$ configuration on the basis of the pure seniority $B(E2)$ trends. Similar behavior has been predicted for the isomers in neighboring $N=120,122,124$ isotonic chains~\cite{ressler2005}. However, the recent $B(E2)$ measurements of the $8^+$ isomeric state in $^{214,216}$Th do not follow pure seniority estimates and require the mixing of $f_{7/2}$ orbital~\cite{zhou2021,zhang2019}, complicating their corresponding configurations. Hence, more investigation into the nucleonic configurations for these ${9/2}^-, {8}^+$, and ${21/2}^-$ isotonic isomers is required.  

Since the moments are more sensitive to nucleonic configurations, it is puzzling that the $Q-$moment trend of ${9/2}^-$ isomers, which is usually asuumed to arise from pure proton-$h_{9/2}$ orbital, does not support the pure seniority (single-j) estimates. Also, the $g-$factors of these ${9/2}^-, {8}^+$, and ${21/2}^-$ isotonic isomers~\cite{stone201419} remain quite far from the pure Schmidt proton-$h_{9/2}$ value. Efforts in the past have been made to explain this using particle-hole excitations, core-polarization effects and meson exchange currents etc.~\cite{towner1977,stuchbery1993,khuyagbaatar2005}. However, a consistent understanding of the decay properties and moments of these isotonic isomers still remains an open puzzle. In view of our recent efforts using generalized seniority~\cite{maheshwari2016,maheshwari20161,maheshwari2019,maheshwari20191}, it would be timely to explore this puzzle by investigating the nucleonic configurations in these isotonic isomers and search for a consistent multi-j configuration to explain both their decays and moments. 

In the present paper, we investigate the ${9/2}^-, {8}^+$, and ${21/2}^-$ isotonic isomers in $N=126$ closed shell and compare with those in $N=124$ (two neutron holes) and $N=128$ (two neutron particles) chains. We use the generalized seniority approach to obtain a consistent understanding of the decay properties as well as the moments. The generalized seniority suggested orbital-occupancies have also been validated with the microscopic shell model for $N=126$ chain. We have additionally explored the ${9/2}^-$ odd-proton states in Tl ($Z=81$), Bi ($Z=83$), At ($Z=85$) and Fr ($Z=87$) isotopes to test the symmetry arguments in this mass region.  

\section{Generalized Seniority Scheme}

The concept of seniority $(v)$ emerges out of the pairing correlations in shell model and may simply be defined as the number of unpaired nucleons for a given state~\cite{racah1943,racah1952,flowers1952}. It is usually described by using the quasi-spin algebra of Kerman~\cite{kerman1961} and Helmers~\cite{helmers1961}. The pair creation operator $S_j^+ = \sum_m {(-1)}^{j-m} a_{jm}^+ a_{j,-m}$ and pair annihilation operator $S_j^-$ which is the Hermitian conjugate of $S_j^+$, satisfy the $SU(2)$ Lie algebra~\cite{shalit1963,casten1990,heyde1990,talmi1993}. For multi-j degenerate orbitals, the generalized seniority scheme~\cite{arima1966} should be used by defining a generalized pair creation operator $S^+ = \sum_{j} S^+_j$, where the summation over $j$ takes care of the multi-j situation. 

In this paper, we invoke the generalized seniority (GS) scheme by defining the quasi-spin operators as $S^+ = \sum_j {(-1)}^{l_j} S^+_j$ ~\cite{arvieu1966} with $l_j$ being the orbital angular momentum quantum number. The corresponding pairing Hamiltonian can be defined as $H= 2 S^+ S^- $, with the energy eigen values $[2s(s+1)-\frac{1}{2} (\Omega-n)(\Omega+2-n)]$ $= \frac{1}{2} [(n-v) (2 \Omega+2-n-v)]$. The  total quasi-spin $s=\sum_j s_j$ having generalized seniority $v=\sum_j v_j $ arises from the multi-j $\tilde{j}=j \otimes j' \otimes....$ configuration with pair degeneracy of $\Omega= \sum_j \frac{2j+1}{2}=\frac{2\tilde{j}+1}{2}$. The total number of nucleons $n=\sum_j n_j$ and the generalized seniority $v=\sum_j v_j$ remain an integer within the quasi-particle picture of shared occupancies. The generalized pair-operators $S^+$ and $S^-$ also satisfy quasi-spin $SU(2)$ algebras with generalized seniority as a quantum number. Consequently, the electric transition probabilities exhibit a parabolic behavior as a function of particle number for both odd and even multipole transitions. We recall the generalized seniority selection rules and reduction formulas from our previous works~\cite{maheshwari2016,maheshwari20161,jain2017,jain20171,maheshwari2017,maheshwari2019,maheshwari20191,maheshwari2021,maheshwari2020} in the following expressions for electric multipole $L$ (even or odd) operators:\\
(a)	For generalized seniority preserving ($\Delta v=0$, $v \rightarrow v$) transitions,
\begin{eqnarray}
\langle {\tilde{j}}^n v l J_f ||\sum_i r_i^L Y^{L}(\theta_i,\phi_i)|| {\tilde{j}}^n v l' J_i \rangle = \Bigg[ \frac{\Omega-n}{\Omega-v} \Bigg] \quad \quad \quad \quad && \nonumber \\
\times \langle {\tilde{j}}^v v l J_f ||\sum_i r_i^L Y^{L}(\theta_i,\phi_i)|| {\tilde{j}}^v v l' J_i \rangle && \label{dv0}
\end{eqnarray}
(b)	For generalized seniority changing ($\Delta v=2, v \rightarrow v\pm2$) transitions,
\begin{eqnarray}
\langle {\tilde{j}}^n v l J_f || \sum_i r_i^L Y^{L}(\theta_i,\phi_i)|| {\tilde{j}}^n v\pm 2 l' J_i \rangle \quad \quad \quad && \nonumber \\ \quad \quad \quad \quad = 
  \Bigg[ \sqrt{\frac{(n-v+2)(2\Omega+2-n-v)}{4(\Omega+1-v)}} \Bigg] && \nonumber \\ \quad \quad \quad \quad \quad \times \langle {\tilde{j}}^v v l J_f ||\sum_i r_i^L Y^{L}(\theta_i,\phi_i)|| {\tilde{j}}^v v\pm 2 l' J_i \rangle &&\label{dv2}
\end{eqnarray} 
where $l$ and $l'$ determine the parities of final $J_f$ and initial $J_i$ states. Also, $r^L$ and $Y^L$ are, respectively, the radial and spherical harmonic parts of the electric multipole operator. The reduction formula in Eq. (\ref{dv0}) can be used to calculate the reduced transition probabilties of isomeric states as well as $Q-$moments. We have already touched upon the origin of isomers in various semi-magic isotopes and neighborhood, along with their electromagnetic properties such as reduced transition probabilities, half-lives, $Q-$moments, $g-$factors etc.~\cite{maheshwari2016,maheshwari20161,jain2017,jain20171,maheshwari2019,maheshwari20191}. As a consequence, we established a new kind of seniority isomers decaying via odd-electric tensor which involves a parity change allowed in multi-j environment~\cite{maheshwari2016}. The role of suggested configuration mixing was also found to be crucial in resolving the long-standing puzzle of the double-hump $B(E2)$ trends for the first excited $2^+$ states~\cite{maheshwari20161}.  
This GS scheme for multi-j degenerate orbitals simply works quite well when dominating orbital is surrounded by some lower$-j$ orbitals lying closely in energy. For example in Sn isomers, the dominating and intruder $h_{11/2}$ orbital gets active together with closely lying $d_{3/2}$ and $s_{1/2}$ orbitals~\cite{maheshwari2016}. In other case when dominating orbital is surrounded by quite high$-j$ orbitals and are well spaced in energy, the realistic case of non-degeneracy needs to be tackled. In the present paper, we restrict the occupancy of these higher$-j$ orbitals making them less probable with respect to the dominating orbital. This preserves the algebra while taking care of the less chances of mixing of the neighboring higher$-j$ orbitals $(j_{nei})$ lying far in energy to the dominating orbital $(j_{dom})$ and therefore, handles the non-degeneracy of the multi-j situation. The total occupancy may then be written as $n= A ~ n_{j_{dom}}+ B ~ \sum_{j_{nei}} n_j$; with $A$ and $B$ deciding the contribution of respective occupancies. The corresponding pair degeneracy can be defined as
\begin{eqnarray}
\Omega= A ~\frac{2j_{dom}+1}{2} + B ~ \sum_{j_{nei}} \frac{2j+1}{2}=\frac{2\tilde{j}+1}{2}
\end{eqnarray} 

The nuclear moments provide a good test for the purity of configuration and are most sensitive to the orbitals with unpaired nucleons. 
We have recently proposed a phenomenological model \cite{maheshwari2019} to calculate the $g-$factor trends by merging the idea of generalized seniority with the well-known Schmidt model \cite{schmidt1937} termed as `Generalized Seniority Schmidt Model' (GSSM). The GSSM expressions \cite{maheshwari2019} are obtained by extending the Schmidt model of single-j to the effective multi-j $\tilde{j}=j \otimes j' \otimes....$, and can be written as:
\begin{eqnarray}
g  =& \frac{1}{\tilde{j}} \Bigg[ {\frac{1}{2} g_s}+ (\tilde{j}- \frac{1}{2}) g_l \Bigg]; \tilde{j}=\tilde{l}+\frac{1}{2}\nonumber\\ 
g  =& \frac{1}{\tilde{j}+1} \Bigg[ -\frac{1}{2} g_s + (\tilde{j}+ \frac{3}{2}) g_l \Bigg]; \tilde{j}=\tilde{l}-\frac{1}{2} 
\end{eqnarray}
where $g_s$ and $g_l$ are taken to be 5.59 $n.m$. and 1 $n.m.$ for protons, while $-3.83$ $n.m.$ and 0 $n.m.$ for neutrons, respectively. The GSSM calculated results come closer to the experimental data than the pure Schmidt model (single-j) for various seniority isomers in and around semi-magic isotopes~\cite{maheshwari2019,maheshwari20191}. This may be correlated to the spin quenching of $g_s$ by the amount of $j / \tilde j$ in comparison to the Schmidt model. However, $g_l$ in GSSM remains nearly similar to that in the Schmidt model. The multi-j GSSM takes care of proper configuration mixing for the magnetic moment operator via valence nucleons which takes care of spin quenching as in any first order perturbation theory~\cite{zamick1971,nomura1972}. Further quantitative matching may need the incorporation of higher-order microscopic effects as in other microscopic methods involving core polarisation, meson exchange currents etc.~\cite{towner1987}. Interestingly, GSSM results do not need any kind of tuning to estimate the amount of spin quenching for explaining the experimental data. In GSSM, the spin quenching is governed by the multi-j configuration $(\tilde j)$ as suggested by generalized seniority, which consistently explains other nuclear properties also. 

\section{The ${9/2}^-$, ${8}^+$, and ${21/2}^-$ isotonic isomers}

\begin{figure}[!htb]
\centering
\includegraphics[width=0.5\textwidth,height=8cm]{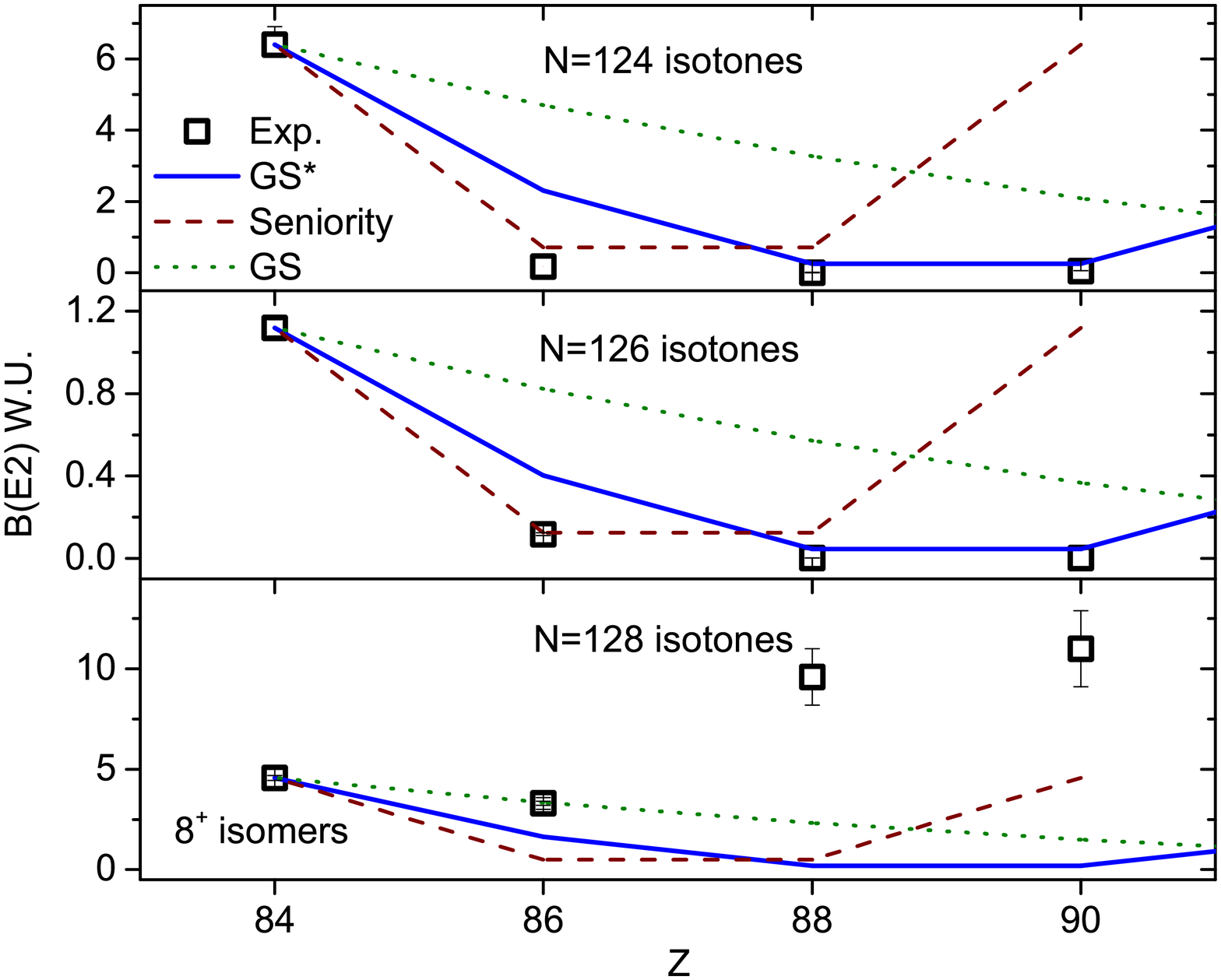}
\caption{\label{fig:be2_8}(Color online) Experimental~\cite{ensdf} and calculated $B(E2)$ trends for the $8^+$ isomers in even$-A$ $N=124$, $N=126$ and $N=128$ isotonic chains. GS$^*$ trends are computed with pair degeneracy $\Omega=7$ suggesting a limited mixing of neighboring $f_{7/2} \otimes i_{13/2}$ orbitals in the $h_{9/2}$ dominated isotonic isomers. Pure seniority and GS calculated trends are also shown for comparison.} 
\end{figure}

\begin{figure}[!htb]
\centering
\includegraphics[width=0.5\textwidth,height=8cm]{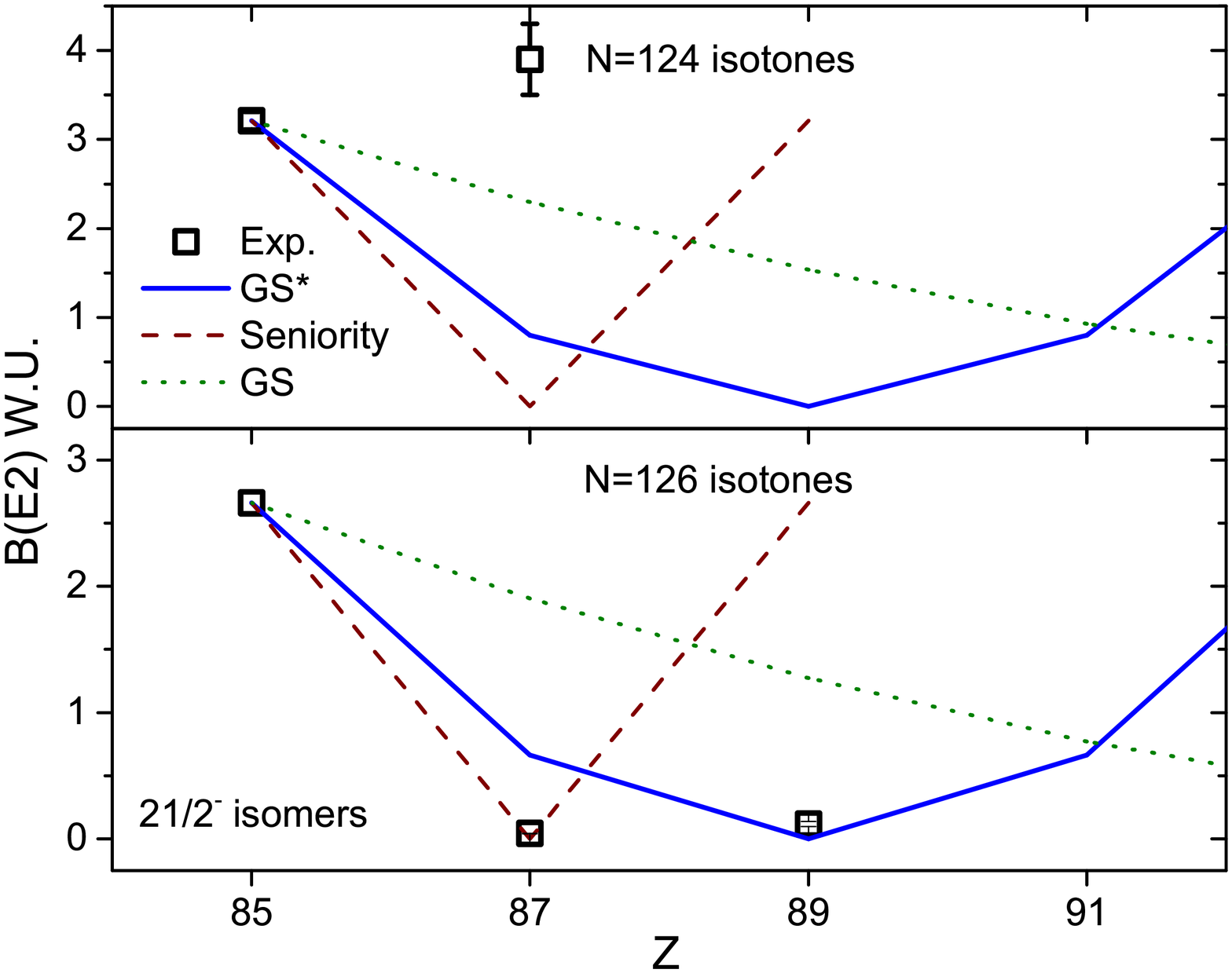}
\caption{\label{fig:be2_21}(Color online) Same as Fig. \ref{fig:be2_8} but for the ${21/2}^-$ isomers in $N=124,126$ isotones.} 
\end{figure}

\begin{figure*}[!htb]
\centering
\includegraphics[width=0.9\textwidth,height=10cm]{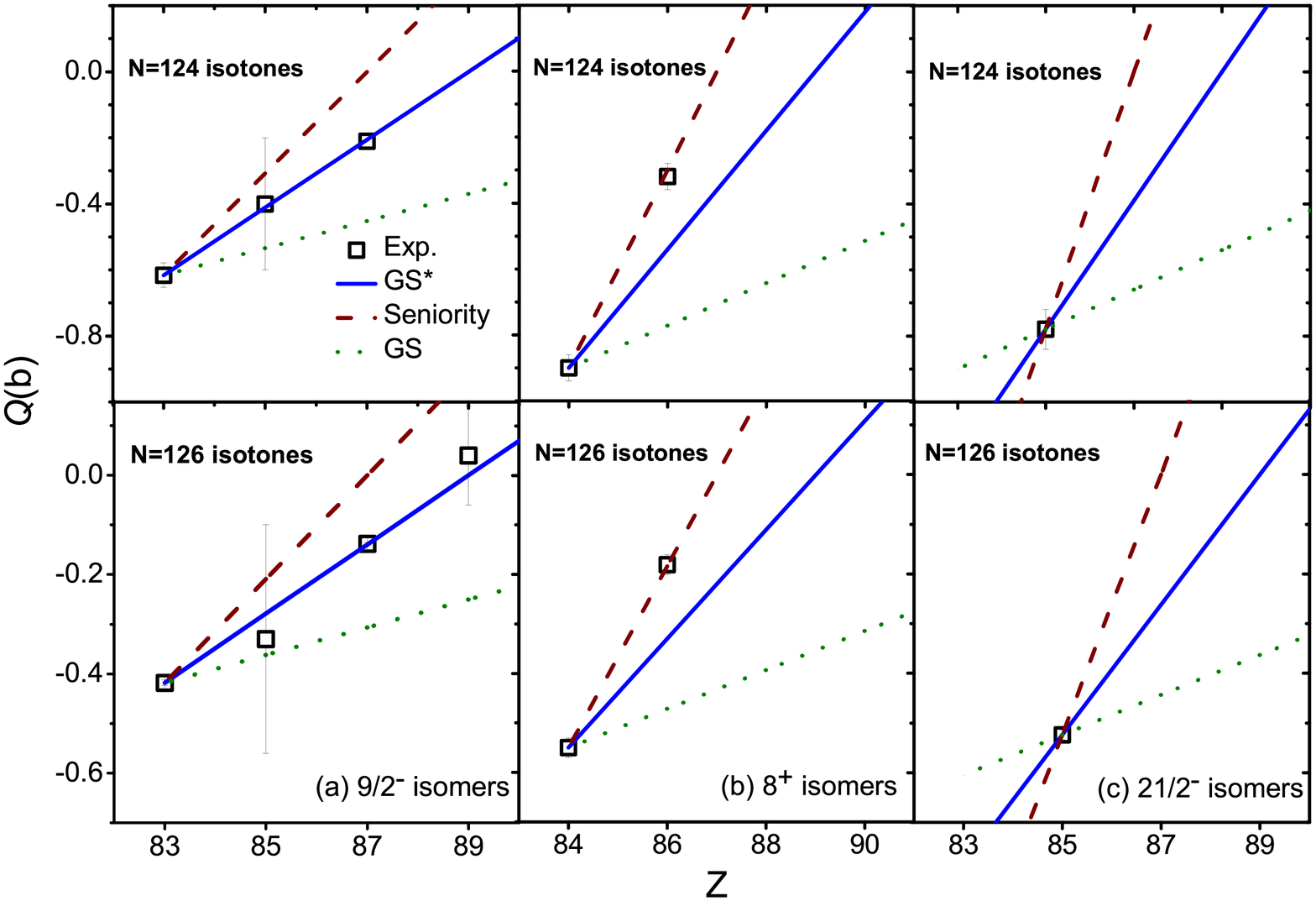}
\vspace{-0.2cm}
\caption{\label{fig:q_isotones}(Color online) Experimental~\cite{stone2021} and calculated $Q-$moment trends for the (a) ${9/2}^-$, (b) $8^+$, and (c) ${21/2}^-$ isotonic isomers in $N=124,126$ chains.} 
\end{figure*}

We present the generalized seniority results for the ${9/2}^-$, ${8}^+$, and ${21/2}^-$ isotonic isomers in $N=124$, $N=126$, and $N=128$ chains. The dominant active orbital for these isomers is proton-$h_{9/2}$ in all the three chains, whereas the mixing of nearby proton-$f_{7/2}$ and proton-$i_{13/2}$ orbitals in the total wave functions of these isomers can not be ruled out. Using the generalized seniority scheme, the total pair degeneracy $\Omega$ turns out to be $16$ corresponding to the \{$h_{9/2} \otimes f_{7/2} \otimes i_{13/2}$\} proton multi-j configuration. The $Z=82$ is taken to be proton-core. Since these ${9/2}^-$, ${8}^+$, ${21/2}^-$ isomers are dominated by $h_{9/2}$ orbital, the mixing of neighboring high$-j$ orbitals in the respective isomeric wave functions can not go beyond a certain limit in realistic cases. To take care of this, we restrict the maximum occupancy of neighboring $f_{7/2} \otimes i_{13/2}$ orbitals to various possible limits by using the definition $n= A ~ n_{j_{dom}}+ B ~ \sum_{j_{nei}} n_j$; with $A$ and $B$ deciding the limits to the respective occupancies of dominant ($j_{dom})$ and neighoring ($j_{nei}$) orbitals. The best and optimum results for both the decays and moments ($Q-$moments and $g-$factors) of these ${9/2}^-$, ${8}^+$, and ${21/2}^-$ isotonic isomers are found with $A=1 $ and $B=0.18$, that is the occupancy of $f_{7/2} \otimes i_{13/2}$ orbitals are taken to be 4, resulting in a total occupancy of 14 on adding the full occupancy of $h_{9/2}$ orbital. The corresponding pair degeneracy becomes $\Omega=7$. This choice of $\Omega$, in a way, takes care of the non-degeneracy of the multi-j environment where $h_{9/2}$ orbital is more probable ($\sim 71 \%$) than the other $f_{7/2} \otimes i_{13/2}$ orbitals ($\sim 29 \%$). These results are labelled by GS$^*$ in $B(E2)$ and $Q-$moment trends, and by GSSM$^*$ in $g-$factor trends. The calculations are done by fitting one of the experimental data for obtaining the $B(E2)$ and $Q-$moment trends. However, no such fitting is required for calculating the $g-$factor trends using GSSM formulas. The multi-j configuration $\tilde{j}$ has been considered to be originating from $\tilde{l}-1/2$ in GSSM calculations since $h_{9/2}$ dominates for the considered isomers in this work.

\subsection{$B(E2)$ rates}

Fig. \ref{fig:be2_8} shows the experimental~\cite{ensdf} and GS$^*$ ($\Omega=7$) calculated $B(E2)$ rates for the yrast $8^+$ isomers in even-A, $N=124,126,$ and $128$ isotonic chains. All the experimental data are listed in Table \ref{tab:data_BE2}. The GS$^*$ calculated $B(E2)$ trends using $v=2$ are able to explain the data very well in all the chains except the two measured values for $Z=88,90$ in $N=128$ chain. This hints towards a change in configuration for these yrast $8^+$, $N=128$ isotonic isomers while moving from $Z=86$ to $Z=88$. This may further be confirmed using their respective moments, if available.

\begin{table}[!htb]
\caption{\label{tab:data_BE2} $B(E2)$ values (in Weisskopf Units, W.U.) for the $8^+$ and ${21/2}^-$ isotonic isomers in $N=124,126,128$ chains. The data have been taken from ENSDF~\cite{ensdf}, unless otherwise stated. The uncertainties are shown in the parentheses. }
\begin{center}
\begin{tabular}{c c c c c}
\hline
Z & $J^\pi$ & $N=124$ & $N=126$ & $N=128$ \\
\hline
84 & $8^+$ & 6.4(5) & 1.12(4)   & 4.56(12) \\
86 & $8^+$ & 0.173$^*$ & 0.117(7)   & 3.3$(^{+3}_{-1})$ \\
88 & $8^+$ & 0.0094$(^{+30}_{-20})$ & 0.00137(17) & 9.6(14) \\
90 & $8^+$ & 0.055(7)$^@$ & 0.000656$^a$ & 11.0(19) \\
\hline
85 & ${21/2}^-$ & 3.21(10)  &  2.66(10)   &   \\
87 & ${21/2}^-$ & 3.99(4)  & 0.0439(20)   &  \\
89 & ${21/2}^-$ &   & 0.119(20)  &  \\
\hline
\multicolumn{5}{l}{$^*$calculated using $E_\gamma$ of 45 keV~\cite{ensdf}} \\
\multicolumn{5}{l}{$^@$Zhou $et$ $al.$~\cite{zhou2021}} \\
\multicolumn{5}{l}{$^a$estimated assuming 26 keV $E_\gamma$ and 126 $\mu s$ half-life~\cite{zhang2019}} 
\end{tabular}
\end{center}
\end{table}

Pure seniority (with $\Omega=5$) and generalized seniority (with $\Omega=16$) trends are also shown for comparison. Pure seniority results are able to explain the data until $Z=88$ in both $N=124$ and $N=126$ isotones. But the role of configuration mixing (as suggested by GS$^*$) becomes crucial while explaining the recently measured $8^+$ isomers in $^{214,216}$Th \cite{zhou2021,zhang2019}, where pure seniority would lead to a large $B(E2)$ value. In contrast, it fits into the GS$^*$ trend leading to a much longer-lived isomer. In 2005, Ressler \textit{et al.}~\cite{ressler2005} have explained the $8^+$ isomers in various $N=120,122,124,126$ isotonic chains using pure seniority scheme. However, no data were available for $Z=90$, Th isotopes at that time. These data are now available and change the picture drastically. Interestingly, the GS results with $\Omega=16$ lie quite far to the data supporting our initial assumption of limited mixing from the neighboring $f_{7/2} \otimes i_{13/2}$ orbitals in these $h_{9/2}$ dominated isomeric states. The $8^+$ isomers also exist in $^{216,218}$U without $E_\gamma$ information and most probably are $\alpha-$decaying~\cite{leppanen2007,ma2015}.  

We further present in Fig.~\ref{fig:be2_21} the experimental and GS$^*$ calculated $B(E2)$ trends for $v=3$, ${21/2}^-$ isomers in odd-A, $N=124$ and $N=126$ chains. In $N=126$ isotones, the measured $B(E2)$ value for Ac, $Z=89$ isotope clearly supports the configuration mixing as suggested by GS$^*$, in contrast to the pure seniority expectations. The GS trends are also shown for comparison and lie quite far from the experimental data. In $N=124$ isotones, the configuration for Fr, $Z=87$ isotope seems to be different than the used proton configuration as it is situated very far from the theoretical trends. No data are available for these $v=3$ isomers in $N=128$ isotones. 

\subsection{$Q-$moments}

\begin{table}[!htb]
\caption{\label{tab:data_Q} $Q-$moment values (in the units of b) for the ${9/2}^-$, $8^+$ and ${21/2}^-$ isotonic isomers in both the $N=124$ and $N=126$ chains. The data have been adopted from Stone's compilation~\cite{stone2021}, unless otherwise stated. The uncertainties are shown in the parentheses. In case of multiple measurements, the weighted average value has been adopted. }
%\begin{center}
\begin{tabular}{c c c c}
\hline
Z & $J^\pi$ & $N=124$ & $N=126$ \\
\hline
83 & ${9/2}^-$ & -0.62(4)$^{*}$ & -0.418(6)$^{*}$ \\
85 & ${9/2}^-$ & -0.4(2) & -0.33(23)$^{@}$ \\
87 & ${9/2}^-$ & -0.21(2) & -0.138(3) \\
89 & ${9/2}^-$ &  & +0.04(10) \\
\hline 
84 & $8^+$ & (-)0.90(4) & [-0.55(2)]$^b$ \\
86 & $8^+$ & \{-\}0.32(4)$^{a}$ & [-0.18(2)]$^b$ \\
\hline
85 & ${21/2}^-$ & (-)0.78(6) & [(-)0.524(10)]$^b$    \\
\hline
\multicolumn{4}{l}{$^*$Skripnikov $et$ $al.$~\cite{skripnikov2021}}  \\
\multicolumn{4}{l}{$^@$Cubiss $et$ $al.$~\cite{cubiss2018}}  \\
\multicolumn{4}{l}{$^a$Sign assumed by us based on the systematics} \\
\multicolumn{4}{l}{$^{b}$not measured, estimated from corresponding $B(E2)$ values}
\end{tabular}
%\end{center}
\end{table}

Fig. \ref{fig:q_isotones}(a) presents a comparison of experimental and GS$^*$ calculated $Q-$moment trends for the ${9/2}^-$ isomers in $N=124$ and $N=126$ chains.  All the experimental data have mostly been adopted from Stone's latest compilation \cite{stone2021} unless otherwise stated and listed in Table \ref{tab:data_Q} with details. No data are available in $N=128$ chain. The GS$^*$ calculations using $v=1$ and $\Omega=7$ explain the data in both $N=124$ and $N=126$ chains very well, whereas pure seniority and GS results lie quite far from the data. This confirms the role of multi-j configuration and configuration mixing as proposed by $\Omega=7$ in GS$^*$ calculations. Similar $Q-$moment trends are shown for the $v=2$, $8^+$ isotonic isomers in Fig. \ref{fig:q_isotones}(b), where data lie on the pure seniority trend. This is due to the fact that these values in $N=126$ chain are not measured but estimated from the respective $B(E2)$ values of $8^+$ isomers using pure $h_{9/2}$ configuration~\cite{dafni1985,mahnke1987}. For $N=124$ isotones, the measured $Q-$values are derived by using the quadrupole coupling constants calculated while studying $N=126$ isotones~\cite{berger1986,mahnke1987}. Direct measurements of $Q-$moments for these $8^+$ isomers will play a crucial role to confirm the configuration mixing, particularly when the $Q-$moments of ${9/2}^-$ isomers are in line with GS$^*$ expectations. There is very limited experimental information  for the $Q-$moments of ${21/2}^-$ isomers available only in At isotope $(Z=85)$ for both $N=124$ and $N=126$ chains. Fig. \ref{fig:q_isotones}(c) shows theory expectations for these $v=3$, ${21/2}^-$ isomers from GS$^*$, pure seniority and generalized seniority. Further measurements may confirm the situation.         

\subsection{$g-$factors}

\begin{figure}
\centering
\includegraphics[width=0.5\textwidth,height=8cm]{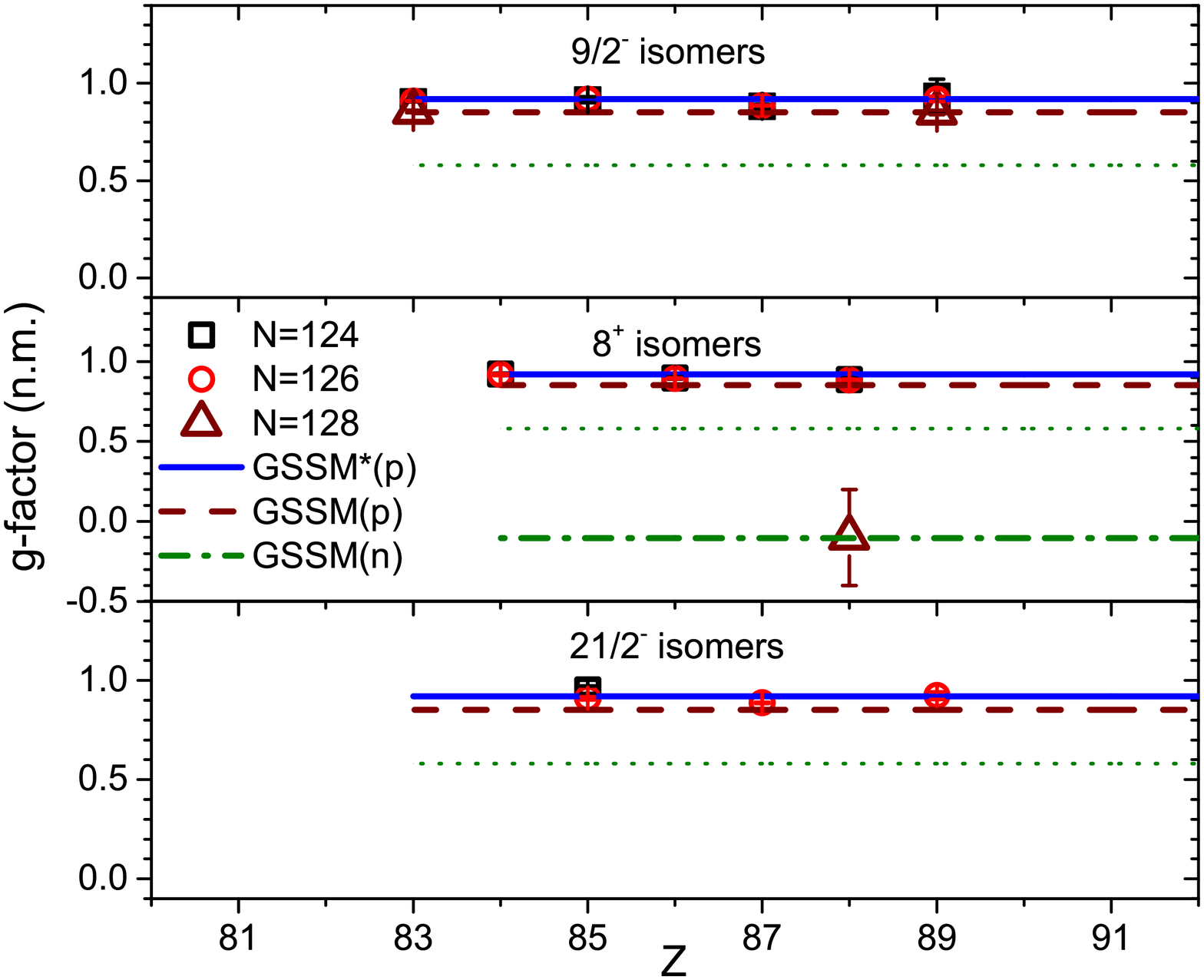}
\caption{\label{fig:g_factor}(Color online) Experimental~\cite{stone201419} and calculated $g-$factor trends for the ${9/2}^-$, $8^+$ and ${21/2}^-$ isomers in all the three $N=124$, $N=126$ and $N=128$ isotonic chains. GSSM$^*$(p) refers to the values obtained using the proton-configuration mixing corresponding to $\Omega=7$, as adopted in $B(E2)$ and $Q-$moment trends. The dotted line represents the pure Schmidt-moment value of $g-$factor in proton $h_{9/2}$ orbital.} 
\end{figure}

We now present in Fig.~\ref{fig:g_factor}, the $g-$factor trends of ${9/2}^-$, $8^+$ and ${21/2}^-$ isomers in all the three $N=124$, $N=126$ and $N=128$ isotonic chains. The GSSM$^*$ results are shown using the pair degeneracy of $\Omega=7$ and corresponding configuration mixing. The $h_{9/2}$ orbital is occupied with $71\%$ probability while rest of the orbitals with $29\%$. We can therefore obtain the total g-factor for these ${9/2}^-$, $8^+$ and ${21/2}^-$ isomers using $ 0.84 \times g_{h_{9/2}} +0.54 \times g_{ f_{7/2} \otimes i_{13/2}}$, where $0.84$ and $0.54$ are the respective mixing amplitudes. The calculations are done using $g_{h_{9/2}}$ as $0.58$ (pure Schmidt proton value) and $g_{ f_{7/2} \otimes i_{13/2}}$ as $0.80$ obtained from GSSM formula for protons in $\tilde{l}-\frac{1}{2}$ configuration. This leads to a total g-factor value of $0.92$ (shown as GSSM$^*$(p) in Fig. \ref{fig:g_factor}) which comes very close to the experimental data for all the three ${9/2}^-$, $8^+$ and ${21/2}^-$ isomers arising from proton $ h_{9/2} \otimes f_{7/2} \otimes i_{13/2}$ configuration as $v=1,2$ and 3, respectively. Experimental data have been adopted from Stone's moment compilations \cite{stone201419} unless otherwise stated in Table \ref{tab:data_g}. Pure Schmidt proton value for ${h_{9/2}}$ and GSSM(p) value for ${h_{9/2} \otimes f_{7/2} \otimes i_{13/2}}$ (with $\Omega=16$) are also shown for comparison. 

The calculated values using protons come closer to the experimental data for both $N=124$ and $N=126$ chains. However, the measured $g-$factor is quite different for $^{216}$Ra ($Z=88,N=128$) (sign assumed to be negative on the basis of systematics). This value lies near the GSSM(n) trend calculated using multi-j neutron $g_{9/2} \otimes i_{11/2} \otimes j_{15/2}$ configuration. This is how one can decide the nature of active nucleons in the generation of these $8^+$ isomers, which can further be related to the deviation in Fig. \ref{fig:be2_8} for $N=128$ isotones. Future moment measurement for the $8^+$ isomers in Th isotopes and ${21/2}^-$ isomers in Bi isotopes would be important to test the role of configuration mixing as obtained from GSSM$^*$(p).

\begin{table}[!htb]
\caption{\label{tab:data_g} $g-$factor values obtained from the magnetic moment data (in the units of n.m.) for the ${9/2}^-$, $8^+$ and ${21/2}^-$ isotonic isomers in all the three $N=124,126,$ and $128$ chains. The data have been adopted from Stone's compilations~\cite{stone201419}, unless otherwise stated.}
\begin{center}
\begin{tabular}{c c c c c}
\hline
Z & $J^\pi$ & $N=124$ & $N=126$  & $N=128$\\
\hline
83 & ${9/2}^-$ &  \{+\}0.9051(4)$^*$ & +0.9093(4) & \{+\}0.856(6)$^*$ \\
85 & ${9/2}^-$ & +0.916(15) & +0.92$^@$ &  \\
87 & ${9/2}^-$ & +0.882(11) & +0.887(2) &  \\
89 & ${9/2}^-$ & +0.93(9) & +0.920(13)	& +0.851(11) \\
\hline 
84 & $8^+$ & +0.921(6)  &  +0.919(6) &  \\
86 & $8^+$ & \{+\}0.898(7)$^*$ &  +0.894(2) &  \\
88 & $8^+$ & \{+\}0.887(9)$^*$  & \{+\}0.885(4)$^*$ & \{-\}0.1(3)$^{*,a}$ \\
\hline
85 & ${21/2}^-$ & +0.952(19) & +0.910(8) &  \\
87 & ${21/2}^-$ &  & \{+\}0.887(3)$^*$ &  \\
89 & ${21/2}^-$ &  & \{+\}0.923(19)$^*$ &  \\
\hline
\multicolumn{5}{l}{$^{*}$sign assumed by us based on the systematics} \\
\multicolumn{5}{l}{$^{@}$estimated value}\\
\multicolumn{5}{l}{$^{a}$average $g-$factor value~\cite{schramm1990}}
\end{tabular}
\end{center}
\end{table}

\begin{figure}
\centering
\includegraphics[width=0.52\textwidth,height=8.5cm]{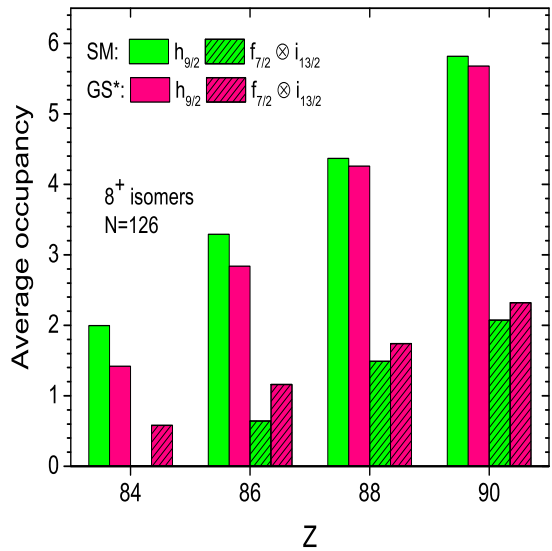}
\vspace{-1.0cm}
\caption{\label{fig:smeven}(Color online) Comparison of shell model and GS$^*$ occupancies for the $8^+$ isomers in even-A $N=126$ isotones.} 
\end{figure}
\section{Shell model occupancies}

\begin{figure}
\centering
\includegraphics[width=0.52\textwidth,height=8.5cm]{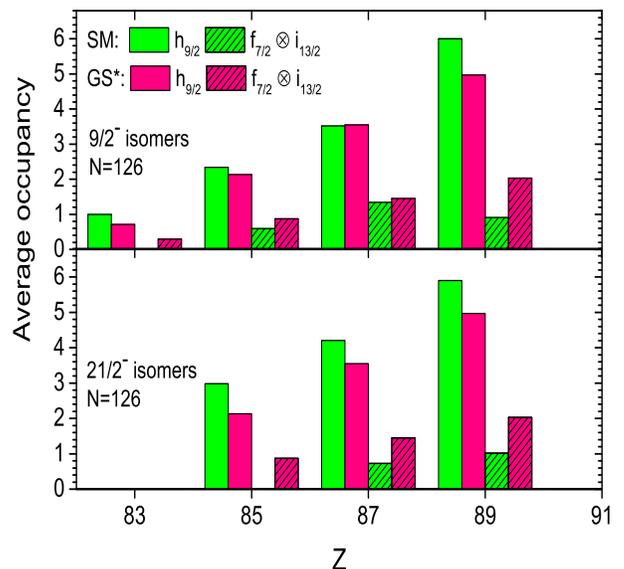}
\vspace{-1.0cm}
\caption{\label{fig:smodd}(Color online) Comparison of shell model and GS$^*$ occupancies for the ${9/2}^-$ and ${21/2}^-$ isomers in odd-A $N=126$ isotones.} 
\end{figure}

We have further tested the generalized seniority (GS$^*$) suggested wave functions with the help of microscopic shell model using the realistic effective Kuo-Herling particle interaction~\cite{warburton1991,mcgrory1975,kuo1971} above $^{208}$Pb for $N=126$ isotones which is shown to work well in explaining the spectroscopic properties of this mass region. The active valence space consists of proton $h_{9/2}$, $f_{7/2}$, $f_{5/2}$, $p_{3/2}$, $p_{1/2}$ and $i_{13/2}$ orbitals having the respective single-particle energies of $-3.799$, $-2.902$, $-0.977$, $-0.681$, $-0.166$ and $-2.191$ MeV.
The shell model Hamiltonian has been diagonalized using NuShellX of Brown and Rae~\cite{brown}. 
The full-space calculations have been done until $Z=88$. Truncations for $^{215}$Ac and $^{216}$Th have been imposed by restricting 4 particles in proton $h_{9/2}$ to meet our computational limitations and allowing the remaining particles for configuration mixing in the chosen proton valence space. We have analyzed the shell model average occupancies for $h_{9/2}$ and $f_{7/2} \otimes i_{13/2}$ orbitals in the even-A and odd-A $N=126$ isotones and compared them in Figs. \ref{fig:smeven} and \ref{fig:smodd} with the GS$^*$ results associated with $\Omega=7$. The agreement between the shell model and GS$^*$ occupancies is quite encouraging, and clearly supports the configuration mixing used by us for these isotonic isomers. Similar results may be expected for $N=128$ and $N=130$ isotonic chains where the dimensions of shell model Hamiltonian would be larger.  

\section{$g-$factors of ${9/2}^-$ states in various isotopic chains}

\begin{figure}[!htb]
\includegraphics[width=0.5\textwidth,height=8cm]{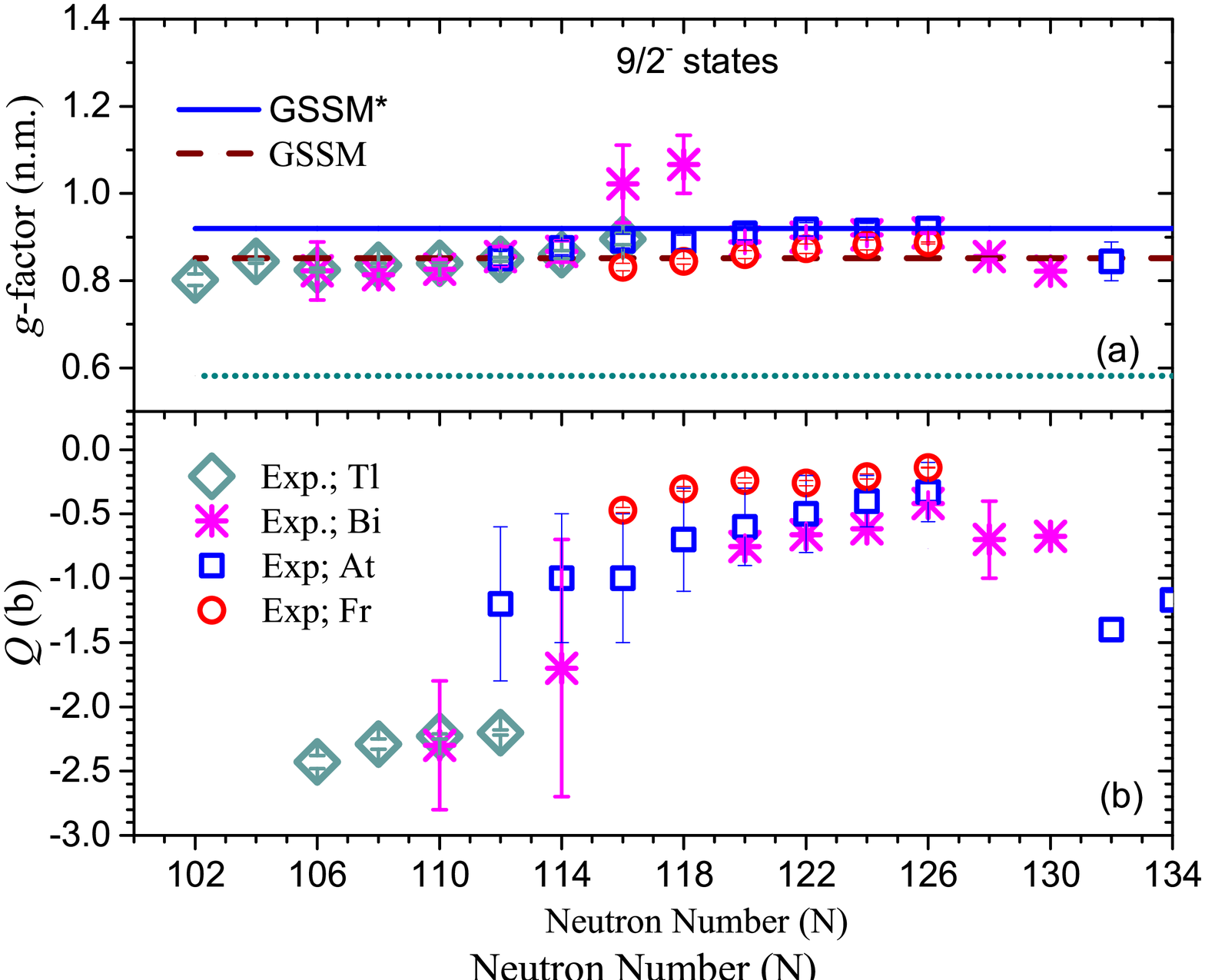}
\caption{\label{fig:g9-q9}(Color online)  Comparison of (a) experimental~\cite{stone201419} and calculated $g-$factor trends, (b) experimental~\cite{stone2021} $Q-$moment trends for the ${9/2}^-$ states in Tl, Bi, At and Fr isotopes. Experimental data for $Q-$moments in Bi isotopes have also been updated from~\cite{skripnikov2021}. The dotted line in panel (a) represents the pure Schmidt-moment value of $g-$factor in proton $h_{9/2}$ orbital.} 
\end{figure}

The $g-$factor values of an odd$-Z$ nucleus can reflect the orbital occupied by the unpaired proton, provided that the nucleus can be described with a rather pure single-particle wave function. Fig.~\ref{fig:g9-q9}(a) presents the experimental~\cite{stone201419} and calculated $g-$factor trends for the proton ${9/2}^-$ states with respect to neutron number in Bi, At and Fr isotopes. In case of multiple measurements, the weighted average value has been adopted. The signs of $g-$factors wherever not known, are assumed to be positive on the basis of systematics. The GSSM and GSSM$^*$ trends, corresponding to $\Omega=16$ and $\Omega=7$ respectively, are shown for comparison. The calculated GSSM and GSSM$^*$ estimates come closer to the experimental data in comparison to the pure Schmidt line for proton $h_{9/2}$ orbital. Further empirical matching may require the core polarisation and particle-hole excitations etc. as discussed by Poppelier and Glaudemans~\cite{poppelier1988}. The $g-$factors of the ${9/2}^-$ isomers in the neutron-deficient Tl isotopes ($Z=81$, one proton-hole configuration) are also shown and found to be in good agreement with those of the ${9/2}^-$ ground states in all the three Bi, At and Fr isotopic chains except for $^{199,201}$Bi. The g-factors support the suggested proton-configuration mixing very well otherwise they would be negative in sign if neutron configurations were to dominate. 

Fig.~\ref{fig:g9-q9}(b) presents the experimental \cite{stone2021} $Q-$moment trends for the ${9/2}^-$ states in Tl, Bi, At and Fr isotopes which exhibit a nearly constant behavior with increasing neutron number (as expected from theory) as the wave functions are mainly dominated by odd-protons. 
All the three Bi, At and Fr isotopes shift towards near spherical value at $N=126$, closed shell configuration. The $Q-$moment trends for these states in $N=120,122$ isotones also support the generalized seniority interpretation, as it varies from negative value to near zero on increasing $Z$, similar to the case discussed in Fig.~\ref{fig:q_isotones}(a) for $N=124,126$ isotones. Future measurements of $Q-$moments for $^{199,201}$Bi may also support the same behavior since the other two known values for $N=116,118$ isotones are in line with this. The difference in the absolute $Q-$moments can be understood due to proton-hole/particle situation so that these ${9/2}^-$ states become excited states for Tl isotopes while they lie as ground states for Bi, At and Fr isotopes. The ${9/2}^-$ isomers in Tl isotopes support oblate deformation along with large negative $Q-$moments. These moments for the ${9/2}^-$ states arising from an odd$-Z$ configuration reflect the role of seniority and associated symmetries due to pairing correlations in Tl, Bi, At and Fr isotopes. This can further be related to the role of proton spectator in the origin of generalized seniority ${13/2}^+, {12}^+$ and ${33/2}^+$ isomers in Hg (with two proton-holes) and Po (with two proton-particles) isotopes so that the isomeric decay properties remain very similar to the case in Pb isotopes~\cite{maheshwari2021}.

\section{Conclusion}
The ${9/2}^-$, $8^+$ and ${21/2}^-$ isomers in and around $N=126$ closed shell are usually understood as arising from pure $h_{9/2}$ configuration on the basis of their $B(E2)$ trends. Recent $B(E2)$ measurements on $^{214,216}$Th isotopes lead to a contrasting behavior and support a very low value instead of pure seniority predictions. In addition, the measured moments of these ${9/2}^-$, $8^+$ and ${21/2}^-$ isomers, which are very sensitive to the nucleonic configurations, lie quite far from the pure $h_{9/2}$-moments. The puzzle of finding a consistent configuration to explain both their decays and moments turns out to be challenging. In this paper, this puzzle has successfully been resolved by using the generalized seniority. These ${9/2}^-$, $8^+$ and ${21/2}^-$ isomers have been established as generalized seniority $v=1$, $v=2$ and $v=3$ isomers using the proton $h_{9/2} \otimes f_{7/2} \otimes i_{13/2}$ multi-j configuration. The best calculated results to explain their experimental $B(E2)$ rates along with the $Q-$moment and $g-$factor trends are found with configuration mixing corresponding to $\Omega=7$. This choice is based on the limited mixing of $f_{7/2} \otimes i_{13/2}$ orbitals in the total wave functions dominated by $h_{9/2}$ orbital. This in a way supports the non-degeneracy of the multi-j orbitals in the quasi-spin scheme. The microscopic shell model calculations for the $N=126$ isotonic isomers support the generalized seniority results. The $g-$factor and $Q-$moment trends for the odd-proton ${9/2}^-$ states in Tl ($Z=81$), Bi ($Z=83$), At ($Z=85$) and Fr ($Z=87$) isotopes are also discussed. Predictions have been made at various places for the gaps in measurements. Such phenomenological model calculations become important especially when the other microscopic model calculations become involved. To conclude, the regular occurrence of isotonic isomeric states is due to the dominance of spherical symmetries which consistently explains their decay properties and moments.

\section*{Acknowledgements}

BM would like to acknowledge the financial support in the form of a Institute Post-doctoral Fellow at IIT Ropar. DC acknowledges the support and facilities received from IIT Ropar to complete this work. AKJ thank Amity University for providing the support and facilities to carry out this work. One of us (AKJ) acknowledges the financial support received from S.E.R.B. (Govt. of India) in the form of a research grant at the Amity University.

\end{document}